\documentclass[a4paper,aps,pre,floatfix,nofootinbib,twocolumn,showpacs]{revtex4-1}

\usepackage[pdftex]{graphicx}
\usepackage{latexsym,amsmath}
\usepackage[pdftex]{color}
\usepackage{hyperref}
\usepackage{url}

\DeclareRobustCommand{\stirling}{\genfrac\{\}{0pt}{}}
\newcommand{\av}[1]{\langle #1 \rangle}

\newcommand{\eps}{\varepsilon}
\newcommand{\lT}{\lambda}
\newcommand{\g}{\gamma}
\newcommand{\kp}{\kappa}
\newcommand{\e}{\epsilon}

\newcommand{\FigPath}{./}

\begin{document}

\title{Temporal percolation in activity driven networks}

\author{Michele Starnini}

\affiliation{Departament de F\'\i sica i Enginyeria Nuclear,
  Universitat Polit\`ecnica de Catalunya, Campus Nord B4, 08034
  Barcelona, Spain}

\author{Romualdo Pastor-Satorras}

\affiliation{Departament de F\'\i sica i Enginyeria Nuclear,
  Universitat Polit\`ecnica de Catalunya, Campus Nord B4, 08034
  Barcelona, Spain}

\date{\today}

\begin{abstract}
  We study the temporal percolation properties of temporal networks by
  taking as a representative example the recently proposed activity
  driven network model [N. Perra \textit{et al.},
  Sci. Rep. \textbf{2}, 469 (2012)]. Building upon an analytical
  framework based on a mapping to hidden variables networks, we
  provide expressions for the percolation time $T_p$ marking the onset
  of a giant connected component in the integrated network. In
  particular, we consider both the generating function formalism,
  valid for degree uncorrelated networks, and the general case of
  networks with degree correlations.  We discuss the different limits
  of the two approach, indicating the parameter regions where the
  correlated threshold collapses onto the uncorrelated case. Our
  analytical prediction are confirmed by numerical simulations of the
  model.  The temporal percolation concept can be fruitfully applied
  to study epidemic spreading on temporal networks. We show in
  particular how the susceptible-infected-removed model on an activity
  driven network can be mapped to the percolation problem up to a time
  given by the spreading rate of the epidemic process. This mapping
  allows to obtain addition information on this process, not available
  for previous approaches.
\end{abstract}


\maketitle

\section{Introduction}

The traditional reductionist approach of network science
\cite{barabasi2005taming} to the study of complex interacting systems
is based in a mapping to a static network or graph, in which nodes
represent interacting units and edges, standing for pairwise
interactions, are fixed and never change in time. This approach has
been proven very powerful, providing a unified framework to understand
the structure and function of networked systems \cite{Newman2010} and
to unravel the coupling of a complex topology with dynamical processes
developing on top of it
\cite{dorogovtsev07:_critic_phenom,barratbook}.  Many networks,
however, are not static, but have instead an evolving topology, with
connections appearing and disappearing with some characteristic time
scales. A static approximation is still valid when such time scales
are sufficiently large, such as in the case of the Internet
\cite{romuvespibook}. In other cases, however, this approximation is
incorrect. This is particularly evident in the case in social
interactions networks, which are formed by a sequence of contact or
communication events, lasting a certain amount of time, which are
constantly being created and terminated between pairs of individuals.
The recent availability of large amounts of data on social networks,
such as mobile phone communications \cite{Onnela:2007}, face-to-face
social interactions~\cite{10.1371/journal.pone.0011596} or large
scientific collaboration databases \cite{2012arXiv1203.5351P} has
spurred the interest in the temporal dimension of social networks,
leading to the development of new tools and concepts embodied in the
new theory of temporal networks \cite{Holme:2011fk}. Key results of
these efforts have been the observation of the ``bursty'',
heterogeneous patterns of social contacts, revealed by distributions
of the time of contact between pairs of individuals, the total time of
contact for an individual, or the gap times between consecutive
interactions involving the same individual, showing a heavy tailed
form
\cite{Oliveira:2005fk,Onnela:2007,Hui:2005,PhysRevE.71.046119,Tang:2010,10.1371/journal.pone.0011596},
or the important effects that the alternation of available edges, and
their rate of appearance, has on dynamical processes running on top of
temporal networks
\cite{Isella:2011,Stehle:2011nx,Karsai:2011,Miritello:2011,dynnetkaski2011,Parshani:2010,Bajardi:2011,Starnini201389,PhysRevLett.98.158702,hoffmann_generalized_2012}.

While the exact characterization of a temporal network is given by the
full ordered sequence of contacts (edges) present in it at time $t$
\cite{Holme:2011fk}, this information is not always easily
available. Thus, sometimes one only has access to coarse-grained
information, in terms of a integrated network, constructed by
integrating the temporal information up to a time $T$, in such a way
that we consider the existence of an edge between nodes $i$ and $j$ in
the integrated version if there was ever an edge in the contact
sequence at any time $t\leq T$.  While these integrated
representations for a fixed (in general large) time have been since
long a useful device to understand the properties of social networks
\cite{newmancitations01}, less in known about the effect of the
integration time $T$ in the structural properties of the integrated
network, an issue which has been recently shown to have relevant
consequences for dynamical processes
\cite{ribeiro_quantifying_2013}. In this context, a particularly
important piece of information is the connectivity properties of the
integrated network as a function of time, and in particular the birth
and evolution of a giant component. Indeed, at a given instant of time
$t$, a temporal network can be represented by a single network
snapshot, which is usually very sparse, composed by isolated edges,
stars or cliques. As we integrate more and more of those snapshots,
the integrated network will grow, until at some time $T_p$ it will
percolate, i.e. it will possess a giant connected component with a
size proportional to the total number of individuals in the
network. The time of the first appearance of this giant component is
not only an important topological property of integrated networks, but
it is also relevant for the evolution of dynamical processes, in the
sense that any process with a characteristic lifetime $\tau < T_p$
will be unable to explore a sizable fraction of the network.

Here we will consider the temporal percolation properties of the
integrated form of temporal networks.  In order to be able to perform
an analytical study, we will focus on the recently proposed activity
driven network model \cite{2012arXiv1203.5351P}, a temporal network
model built on the observation that the establishment of social
interactions is driven by the activity of individuals, urging them,
with different levels of intensity, to interact with their peers.
Building on the mapping of the integrated network ensuing from this
model into the class of network models with hidden variables
\cite{PhysRevE.68.036112,starnini_topological_2013}, we compute
analytic expressions for the percolation time and the size of the
giant component of the integrated network. An added value of our
approach is the possibility to extend the mapping of epidemic
spreading into percolation processes in static networks
\cite{PhysRevE.66.016128} into the temporal case. Thus our results can
be extended to provide the epidemic threshold and the outbreak size of
the susceptible-infected-susceptible epidemic model \cite{anderson92}.

\section{The activity driven model: Definition and topological
  properties}
\label{sec:activ-driv-model}

The activity driven network model \cite{2012arXiv1203.5351P} is
defined in terms of $N$ individuals (agents), each one endowed with an
activity $a_i$, defined as the probability that she starts a social
interaction with other agents per unit time. The activity of the
agents is a random variable, extracted from the activity potential
distribution $F(a)$. Focusing on the emergence of the integrated
network, dynamical creation of links follows an asynchronous scheme
\cite{starnini_topological_2013}: Every time step $\Delta t = 1/N$, an
agent $i$ is chosen uniformly at random. With probability $a_i$, the
agent becomes active and generates $m$ links that are connected to $m$
randomly chosen agents. Those links last for a period of time $\Delta
t$ (i.e. are erased at the next time step). Finally, time is updated
$t \to t + \Delta t$.

The topological properties of the integrated activity driven network
at time $T$ have been studied in Ref.~\cite{starnini_topological_2013}
by means of a mapping to a hidden variables network model
\cite{PhysRevE.68.036112}, which is based on the connection
probability $\Pi_T(a, a')$ that two nodes with activity $a$ and $a'$
are connected in the integrated network at time $T$. For large $N$,
his connection probability takes the form
\cite{starnini_topological_2013}
\begin{equation}
  \label{eq:1}
  \Pi_T(a,a') = 1- \exp\left[ - \lT (a +a') \right],
\end{equation}
where $\lT = T/N$ and we have set $m=1$ to simplify calculations. The
topological properties of the integrated network are encoded in the
propagator $g_T(k|a)$, defined as the probability that a node with
activity $a$ has integrated degree $k$ at time $T$, and whose
generating function $ \hat{g}_T(z|a) = \sum_k g_T(k|a) z^k$ satisfies
the general equation
\cite{PhysRevE.68.036112,starnini_topological_2013}
\begin{equation}
  \ln   \hat{g}_T(z|a) = N \sum_{a'} F(a') \ln \left[ 1-(1-z) \Pi_T(a,a')
  \right].
  \label{eq:3} 
\end{equation}
From the propagator, the degree distribution of the integrated network
at time $T$ is trivially given by
\begin{equation}
  P_T(k)=\sum_a F(a) g_T(k|a).
  \label{eq:pk}
\end{equation}
In the limit of small $\lT$ or $N \gg T$ with constant $T$, which we
assume from now on, the connection probability Eq. (\ref{eq:1}) can be
approximated as
\begin{equation}
  \label{eq:conn_prob}
  \Pi_T(a,a') \simeq \lT (a+a').
\end{equation}
In this same limit $\lT \to 0$, Eq.~(\ref{eq:3}) can be solved,
leading to a  propagator with the form of a Poisson distribution
with mean $T(a+\av{a})$. From it, we obtain an asymptotic degree
distribution \cite{starnini_topological_2013}
\begin{equation}
  P_T(k) \simeq \frac{1}{T} F\left( \frac{k}{T} - 
    \av{a} \right).
\end{equation}
Other topological properties, concerning degree correlations and
clustering spectrum, have been described in
Ref.~\cite{starnini_topological_2013}.

\section{Generating function approach to percolation}
\label{sec:gener-funct-appr}

Percolation in random networks can be studied applying the generating
function approach developed in Ref.~\cite{PhysRevE.64.026118}, which
is valid assuming the networks are degree uncorrelated.  Let us define
$G_0(z)$ and $G_1(z)$ as the degree distribution and the excess degree
distribution (at time $T$) generating functions, respectively, given
by \cite{Newman2010}
\begin{equation}
  G_0(z) = \sum_k P_T(k) z^k, \quad G_1(z) =\frac{G'_0(z)}{G'_0(1)}.  
  \label{eq:generdefs}
\end{equation}
The size of the giant connected component, $S$, is then given by
\begin{equation}
\label{eq:S}
  S = 1 - G_0(u),
\end{equation}
where $u$, the probability that a randomly chosen vertex is not
connected to the giant component, satisfies the self-consistent
equation
\begin{equation}
  \label{eq:S_self}
  u = G_1(u).
\end{equation}
The position of the percolation threshold can be simply obtained by
considering that $u=1$ is always a solution of Eq.~(\ref{eq:S_self}),
corresponding to the lack of giant component. A physical solution
$u<1$, corresponding to a macroscopic giant component, can only take
place whenever $G'_1(1)>1$, which leads to the Molloy-Reed criterion
\cite{molloy95}:
\begin{equation}
  \label{eq:26}
  \frac{\av{k^2}_T}{\av{k}_T} > 2,
\end{equation}
where $\av{k^n}_T = \sum_k k^n P_T(k)$ is the $n$th moment of the
degree distribution at time $T$. 

In the case of the activity driven model, these moments can be
computed noticing, from Eq.~(\ref{eq:pk}), that $\av{k^n}_T = \sum_a
F(a) \sum_k k^n g_T(k|a)$.  Since the propagator has the form of a
Poisson distribution, the moments of the degree distribution simply
read as
\begin{equation}
  \label{eq:moments}
  \av{k^n}_T = \sum_{m=1}^n \stirling{n}{m}  T^m \kp_m,
\end{equation}
where $\stirling{n}{m} $ are the Stirling numbers of the second kind
\cite{gradshteyn2007} and
\begin{equation}
  \kp_m = \sum_a F(a)(a+\av{a})^m = \sum_{i=0}^m \binom{m}{i} \av{a^i}
  \av{a}^{m-i}.
\label{eq:4}
\end{equation}

The ratio $\av{k^2}_T / \av{k}_T $ is a monotonic, growing function of
$T$, and it will fulfil the condition Eq.~(\ref{eq:26}) for $T>T_p^0$,
defining a percolation time
\begin{equation}
  \label{eq:Tc_uncor}
  T_p^0 =  \frac{\kp_1}{\kp_2} =  \frac{2 \av{a}}{\av{a^2} +
    3\av{a}^2}. 
\end{equation}
This percolation time is independent of $N$, and thus guarantees the
fulfilment of the condition $\lT \ll 1$ assumed in the derivation of
Eq.~(\ref{eq:conn_prob}).  We can obtain information on the size of
the giant component $S$ for $T>T_p^0$ from Eqs.~(\ref{eq:generdefs})
and (\ref{eq:pk}), using the Poisson form of the propagator, which
allows to write the simplified expressions
\begin{eqnarray}
\label{eq:genfun0}
  G_0(u) &=& \sum_a  F(a) e^{-(1-u)T(a+\av{a})}, \\
  G_1(u) &=& \frac{1}{2 \av{a}} \sum_a F(a) [a+\av{a}]
  e^{-(1-u)T(a+\av{a})}.  \label{eq:genfun1}  
\end{eqnarray}
From the self-consistent Eq.~(\ref{eq:S_self}), setting $\delta =
1-u$, and solving at the lowest order in $\delta>0$, we find, close to
the transition,
\begin{equation}
  \delta \simeq \frac{2\kp_1}{\kp_3 T^2} \left( \frac{T-T_p^0}{T_p^0}
  \right), 
\end{equation} 
recovering the Molloy Reed criterion, Eq.~(\ref{eq:Tc_uncor}), for the
onset of the giant component.  Since the derivatives of $G_0(u)$ are
finite, we can obtain the size of the giant component $S$ by expanding
Eq.~(\ref{eq:genfun0}) close to $u =1$,
\begin{eqnarray}
\label{eq:S_appr}
  S & \simeq &1 - G_0(1) + \delta G_0'(1) - \frac{\delta^2}{2}
  G_0''(1) \nonumber \\  
  & = & \frac{2 \kp_1^2 }{ \kp_3 T} \left( \frac{T-T_p^0}{T_p^0} \right) -
  \frac{2 \kp_2 \kp_1^2}{\kp_3^2 T^2} \left( \frac{T-T_p^0}{T_p^0}
  \right)^2 . 
\end{eqnarray}
Since Eq. (\ref{eq:S_appr}) is obtained from a Taylor expansion for
$\delta \ll 1$, we expect it to be valid only close to the percolation
threshold.

In order to check the validity of the analytical results developed
above, we consider the concrete case of two different forms of
activity distribution, namely a uniform activity distribution $F(a) =
a_{\textrm{max}}^{-1}$, with $a \in [0, a_{\textrm{max}}]$, and the
empirically observed case of power law activity distribution in social
networks \cite{2012arXiv1203.5351P}, $F(a) \simeq (\g -1) \e^{\g -1}
a^{-\g}$, $a \in [\e, 1]$, where $\e$ is the minimum activity in the
system.  In this last case, we note that the analytical form of the
activity distribution is valid for small $\e$ only in the limit of
large $N$. Indeed, a simple extreme value theory calculation
\cite{gumbel2004statistics} shows that, in a random sample of $N$
values $a_i$, the maximum activity scales as $\min\{ 1, \eps
N^{1/(\gamma-1)} \}$. Therefore, when performing numerical simulations
of the model, one must consider systems sizes with $N > N_c =
\e^{1-\g}$ in order to avoid additional finite-size effects.  In case
of performing simulations for system with small sizes $N<N_c$, for
example to study the finite size effects on the percolation threshold
(see next Section), we used a deterministic power law distribution to
avoid the cutoff effect on the maximum value of the activity due to a
random sampling of values.
 
\begin{figure}[tbp]
  \centerline{\includegraphics[width=8.5cm]{\FigPath/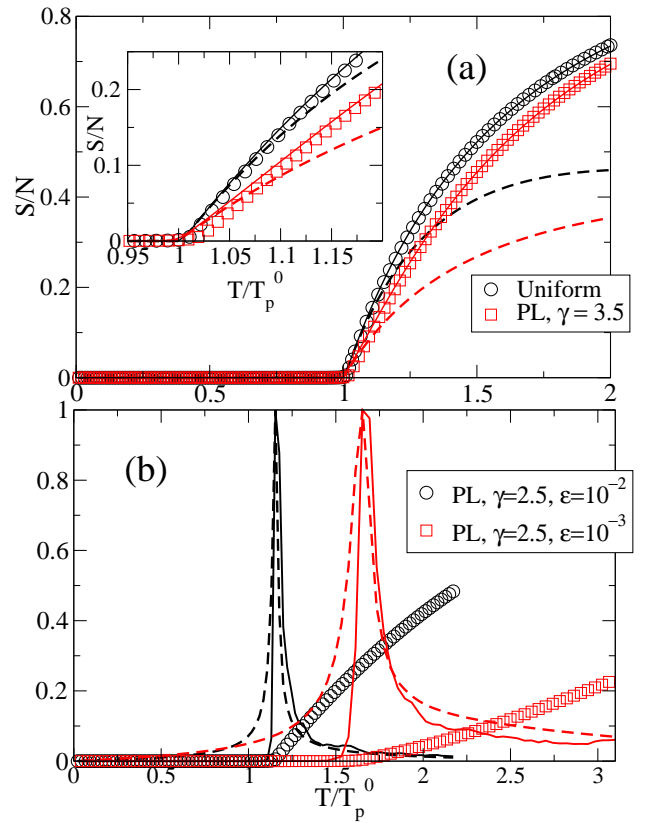}}
  \caption{Rescaled giant component size $S/N$ as a function of the
    rescaled time $T/T_p^0$ for activity driven networks.  (a) Uniform
    ($a_{\textrm{max}} = 0.01$) and power law ($\g = 3.5$, $\e =
    0.01$) activity distributions, compared with the numerical
    integration of the generating function equations (continuous line)
    and the theoretical approximation Eq.~(\ref{eq:S_appr}) (dashed
    line).  (b) Power law activity distribution with $\g = 2.5$, $\e = 10^{-2}$ 
    and $\e = 10^{-3}$.  The peaks of the variance of the giant
    component size, $\sigma(S)^2$, and the susceptibility of the
    clusters size, $\chi(s)$, are plotted in continuous and dashed
    line, respectively.  Results are averaged over $10^2$ runs,
    network size $N=10^7$. }
  \label{fig:S_comp}
\end{figure}

Fig. \ref{fig:S_comp} shows the giant component size $S$ of the
activity driven network as a result of numerical simulations for both
uniform and power law activity distributions.  In
Fig.~\ref{fig:S_comp}(a) we compare $S$ with the analytical
approximation Eq.~(\ref{eq:S_appr}), as well as with the result of a
direct numerical integration of Eqs.~(\ref{eq:S}) and~(\ref{eq:S_self}),
for uniform activity and power-law activity with exponent
$\gamma>3$. In both cases we observe an almost exact match between
numerical simulations and the numerical integration of the generating
function equations, and a very good agreement with the analytical
approximation in the vicinity of the percolation threshold.
Fig.~\ref{fig:S_comp}(b), on the other hand, focuses on power-law
activity distributions with an exponent smaller than three. In this
case we additionally plot a numerical estimation of the percolation
threshold as given by the peak of both the variance of the giant
connected component size $S$, $\sigma(S)^2 = \av{S^2}-\av{S}^2$, and
the susceptibility of the clusters size, $\chi(s) = \sum_{s=2}^{S-1}
s^2 n_s$, where $n_s$ is the number of cluster of size $s$. From this
figure we can see that the numerical percolation threshold strongly
deviates in this case from the theoretical prediction
Eq. (\ref{eq:Tc_uncor}), deviation that increases when the
distribution cutoff $\eps$ becomes smaller.

\section{Effect of degree correlations on the temporal percolation
  threshold}
\label{sec:effect-degr-corr}

The origin of the disagreement for the case of power law activity
distribution with $\gamma < 3$ can be traced back to the effect of
degree correlations in the integrated networks generated by the
activity driven model. Indeed, as stated above, the generating
function technique makes the explicit assumption of lack of degree
correlations \cite{PhysRevE.64.026118}.  However, the integrated
activity driven network has been shown
\cite{starnini_topological_2013} 
to exhibit degree correlations, 
as measured by an average degree of the neighbors 
of the vertices of degree $k$, $\bar{k}^{nn}_T(k)$ \cite{alexei}, being a decreasing function of $k$.
Another, global measure of degree correlations
can be defined in terms of the Pearson correlation coefficient $r$
between the degree of a node and the mean degree of its neighbors
\cite{PhysRevLett.89.208701}, taking the form
\begin{equation}
  \label{eq:general_r}
  r =  \frac{ \av{k} \sum_k k^2 \bar{k}^{nn}(k) P(k) - \av{k^2}^2 }
  {\av{k} \av{k^3} - \av{k^2}^2 }.  
\end{equation} 
We can easily evaluate the sum $\sum_k k^2 \bar{k}^{nn}_T(k) P_T(k) $
by applying the hidden variable formalism presented in
Sec.~\ref{sec:activ-driv-model}. Inserting in Eq. (\ref{eq:general_r})
the first moments of the degree distribution as obtained from
Eq.~(\ref{eq:moments}), the coefficient $r$ in the limit of large $N$ reads
\begin{equation}
  \label{eq:r_th}
  r_T = - \frac{  \left( \sigma_a^2 \right)^2}{  \dfrac{\kp_1\kp_2}{T}
    +  \kp_1\kp_3 -  \kp_2^2 },
\end{equation}
where $\sigma_a^2 = \av{a^2} - \av{a}^2$ is the variance of the activity distribution.  
Both the decreasing functional form of $\bar{k}^{nn}_T(k)$ and the
negative value of $r$ (since $\kp_1\kp_3 > \kp_2^2$ for any
probability distribution with a positive support), indicate the
presence of dissasortative correlations \cite{PhysRevLett.89.208701}
in the integrated activity driven networks, correlations whose
amplitude is modulated by $\sigma_a^2$.

In order to take into account the effect of degree correlations let us
consider the general problem of percolation in a correlated random
network \cite{citeulike:9984067}.  The effect of the degree
correlations are accounted for by the branching matrix 
\begin{equation}
  B_{kk'} =
  (k'-1)P(k'|k),
  \label{eq:9}
\end{equation}
where $P(k'|k)$ is the conditional probability that a node with degree
$k$ is connected to a node with degree $k'$ \cite{alexei}.  The
percolation threshold is determined by the largest eigenvalue
$\Lambda_1$ of the branching matrix $B_{kk'}$ through the condition
$\Lambda_1 = 1$.  If the network is uncorrelated, $\Lambda_1$ reduces
to the ratio of the first two moment of the degree distribution,
$\Lambda_1^0 = \av{k^2} /\av{k} - 1 $, thus recovering the Molloy-Reed
criterion Eq.~(\ref{eq:26}).
In activity driven networks we can compute the largest eigenvalue
$\Lambda_1$ in the limit of small $\lT$ by applying the hidden
variables mapping from Sec.~\ref{sec:activ-driv-model}. In fact, the
conditional probability $P_T(k'|k)$ of the integrated network at time
$T$ can be written as \cite{PhysRevE.68.036112}
\begin{eqnarray}
  P_T(k'|k) &=& \frac{N}{P_T(k)} \sum_{a,a'} g_T(k-1|a') F(a')
  \frac{\Pi_T(a',a)}{\bar{k}_T(a)} \nonumber \\
  &\times& F(a) g_T(k|a),
\end{eqnarray}
where $\bar{k}_T(a) = N \sum_a F(a) \Pi_T(a,a')$
\cite{starnini_topological_2013}. 
From here, the branching matrix takes the form
 \begin{equation}
   B_{kk'} = (k'-1) \left[ p_{k'-1} + \frac{p_{k-1}}{k p_{k}}
    \left( k' p_{k'} - \av{k}  p_{k'-1} \right) \right], 
\end{equation}
where we write $P_T(k)$ as $p_k$ for brevity.
Assuming that the branching matrix is irreducible, and given that it is
non-negative (see Eq.~(\ref{eq:9})) we can compute its largest
eigenvalue by applying Perron-Frobenius theorem \cite{gantmacher} and
looking for a principal eigenvector $v_k$ with positive components. 
Using the ansatz $v_k = 1 + \alpha p_{k-1}/k p_{k}$, we obtain that,
in order to be an eigenvector, the following conditions must be
fulfilled:
\begin{eqnarray}
  \Lambda_1 &=& \av{k}_T  +\alpha \sum_k\frac{(k-1) p_{k-1}^2}{k p_k} \nonumber \\  
  \Lambda_1 \alpha &=& \av{k^2}_T - \av{k}_T -\av{k}_T^2 + \alpha \av{k}_T
  \left(1- \sum_k\frac{(k-1) p_{k-1}^2}{k p_k} \right)\nonumber.  
\end{eqnarray}
One can see that $\sum_k(k-1) p_{k-1}^2/k
p_{k} \simeq 1$, in the limit of large $N$.  Thus we obtain the equation for $\Lambda_1$
\begin{equation}
  \Lambda_1(T)^2 - \av{k}_T \lambda_1(T) - \av{k^2}_T + \av{k}^2_T  +
  \av{k}_T = 0.
  \label{eq:2}
\end{equation}
By using the form of the moments of the degree distribution given by
Eq.~(\ref{eq:moments}), we solve Eq.~(\ref{eq:2}). Excluding the
non-physical solution $\Lambda_1 < 0$, one finally find the largest eigenvalue 
\footnote{From Eq. (\ref{eq:lambda_real}) one can find 
$\alpha = \frac{\sigma_a^2}{(\sqrt{\av{a^2}} + \av{a})} T > 0$, 
confirming the validity of the proposed ansatz for $v_k$.} 
of the branching matrix as
\begin{equation}
  \label{eq:lambda_real}
  \Lambda_1(T) = \left( \sqrt{\av{a^2}} + \av{a} \right) T. 
\end{equation} 
From here, the percolation threshold in activity driven networks
follows as 

\begin{equation}
  \label{eq:perc_real}
  T_p = \frac{1}{\sqrt{\av{a^2}} + \av{a}}.
\end{equation}
 
We can understand the results of Fig.~\ref{fig:S_comp} by comparing
the ratio of the exact threshold $T_p$ with the uncorrelated value $T_p^0$, 
\begin{equation}
  Q = \frac{T_p - T_p^0}{T_p^0} = \frac{\sigma_a^4}{2\av{a} \left(
      \sqrt{\av{a^2} } + \av{a} \right)^3 }. 
  \label{eq:Q}
\end{equation}
In the case of a uniform activity distribution, we have $Q =
13/\sqrt{3} - 15/2 \simeq 5.5 \times 10^{-3}$, and therefore the
temporal percolation threshold is given with very good accuracy by the
uncorrelated expression. For a power-law activity distribution, the
ratio $Q$ depends simultaneously of the exponent $\gamma$ and the
minimum activity $\eps$. Thus, for $\gamma<3$, we have that $Q \sim
\eps^{(\gamma-3)/2}$, which diverges for $\eps\to0$, indicating a
strong departure from the uncorrelated threshold. For $\gamma>3$, on
the other hand, $Q$ becomes independent of $\eps$, and it goes to $0$
in the limit of large $\gamma$. In the case $\gamma=3.5$ and
$\eps=0.01$, for example, we obtain $Q \simeq 1.6 \times 10^{-2}$.
This implies an error of less than $2\%$ in the position of the
percolation threshold as given by the uncorrelated expression,
explaining the good fit observed in Fig.~\ref{fig:S_comp}(a).  

\begin{figure}[tbp]
  \includegraphics[width=8cm]{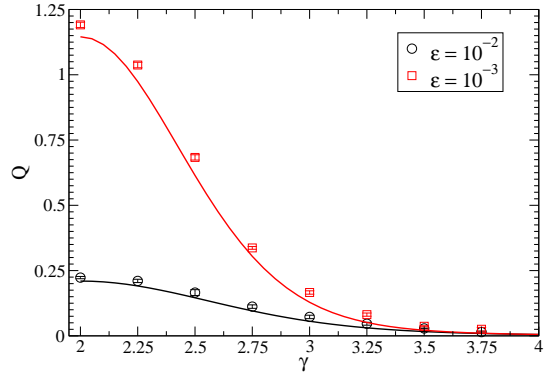}
  \caption{Ratio $Q$, defined in Eq. (\ref{eq:Q}), as a function of
    the exponent $\gamma$ of the activity potential $F(a) \sim
    a^{-\gamma}$, for $\epsilon = 10^{-2}$ and $\epsilon = 10^{-2}$.
    We compare $Q$ as obtained by estimating the percolation threshold
    $T_p$ from the peak of the variance of the giant component size,
    $\sigma(S)^2$, by means of a numerical simulation of a network
    with size $N=10^7$ (symbols), with the prediction of
    Eq. (\ref{eq:Q}) (lines).The results of numerical simulations are averaged over $10^2$ runs.}
  \label{fig:Q}
\end{figure}

\begin{figure}[tb]
  \includegraphics[width=8cm]{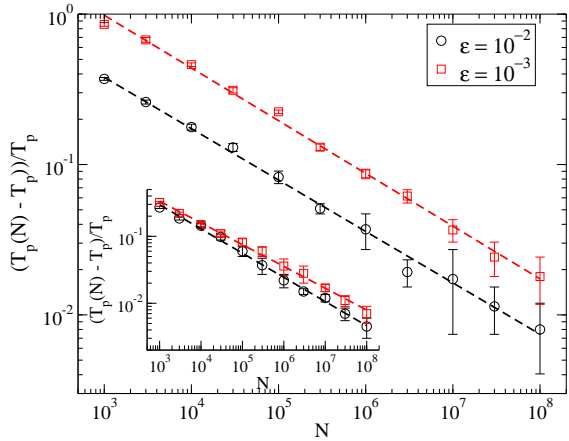}
  \caption{Finite size scaling of the percolation threshold as
    estimated by the peak of the variance of the giant component size,
    $\sigma(S)^2$, for a network with power law activity distribution with
    $\gamma = 2.5$ (main) and $\gamma = 3.5$ for different values of $\epsilon$.
    We plot $(T_p(N) - T_p)/T_p$ as a function of $N$, finding a
    scaling of the form of Eq. \ref{eq:ffs}, plotted in dashed line, 
    with the same exponent $\nu = 0.34 \pm 0.01$ for $\gamma = 2.5$
    and $\nu = 0.34 \pm 0.02$ for $\gamma = 3.5$,
    The results are averaged over $10^2$ runs.
    }
  \label{fig:fss}
\end{figure}

In Fig.~\ref{fig:Q} we show the ratio $Q$ as a function of the
exponent $\gamma$ of a power-law distributed activity potential for
different values of $\epsilon$, computed from numerical simulations by
evaluating the percolation threshold from the peak of the variance of the
giant component size, $\sigma(S)^2$. The numerical result is compared
with the analytical prediction given by Eq. (\ref{eq:Q}).
In this Figure one can see that, although numerical and analytical
results are in quite good agreement, they still do not exactly
coincide for $\gamma < 3$.  This is due to the presence of finite size
effects, which have not been taken into account in the percolation
theory developed.  We can consider the finite size effects on the
percolation time $T_p(N)$ in a network of size $N$ by putting forward
the standard hypothesis of a scaling law of the form
\begin{equation}
\label{eq:ffs}
T_p(N) = T_p + A N^{-\nu}.
\end{equation} 
In Fig. \ref{fig:fss} we plot the rescaled numerical thresholds
$[T_p(N)-T_p]/T_p$ estimated by the peak of the variance of the giant
component size, $\sigma(S)^2$, as a function of the network size $N$.
We can observe that the numerical thresholds $T_p(N)$ asymptotically
tend to the theoretical prediction $T_p$ by following the scaling law
of Eq. (\ref{fig:fss}), with a very similar exponent $\nu \simeq 0.34
\pm 0.02$, and different values of the prefactor $A$ depending on the
values of $\gamma$ and $\epsilon$.  We also checked (data not shown)
that the other method to measure the percolation threshold, through
the susceptibility of the clusters size $\chi(s)$, follows the same
scaling law, with the same exponent $\nu$ and a slightly different
prefactor $A$.

\section{Application to epidemic spreading}
\label{sec:appl-epid-spre}

The concept of temporal percolation can be applied to gain
understanding of epidemic processes on activity driven
temporal networks \cite{2012arXiv1203.5351P}. Let us focus in the
susceptible-infected-susceptible (SIR) model \cite{anderson92}, which
is the simplest model representing a disease that confers immunity and
that is defined as follows: Individuals can be in either of three
states, namely susceptible, infected, and removed. Susceptible individuals acquire the disease by contact with
infected individuals, while infected individuals heal spontaneously
becoming removed, which cannot contract the disease anymore. On a
temporal network, the SIR model is parametrized by the rate $\mu$
(probability per unit time) at which infected individuals become
removed, and by the transmission probability $\beta$ that the
infection is propagated from an infected individual to a susceptible
individuals by means of an instantaneous contact.

We can approach the behavior of the SIR model on activity driven
networks by extending the mapping to percolation developed in
Ref.~\cite{PhysRevE.66.016128} into the temporal case. To do so, let
us consider first a modified SIR model in which individuals stay in
the infected state for a fixed amount of time $\tau$. We define the
transmissibility $\mathcal{T}_{ij}$ as the probability that the
infection is transmitted from infected individual $i$ to susceptible
individual $j$. Considering that contacts last for an amount of time
$\Delta t = 1/N$, the transmissibility can be written as
\begin{equation}
 \mathcal{T}_{ij}(\beta, \tau) = 1 - \left( 1 - \beta  p_{ij} \right)
 ^{\tau N},  
\end{equation}
where $p_{ij} = (a_i + a_j)/N^2$ is the probability that individuals
$i$ and $j$ establish a contact in any given time step $\Delta t$
\cite{starnini_topological_2013}. In the limit of large $N$, we can
thus write
\begin{equation}
 \mathcal{T}_{ij}(\beta, \tau) = 1 - \exp\left(-\frac{\beta \tau
      [a_i+a_j]}{N}\right). 
\end{equation}
From here we can deduce the form of the transmissibility when healing
is not deterministic but a Poisson process with rate $\mu$. In this
case, the probability that an infected individual remains infected a
time $\tau$ is given by the exponential distribution $P(\tau) = \mu
e^{- \mu \tau}$.
Therefore, we can write \cite{PhysRevE.66.016128}
\begin{eqnarray}
\label{eq:transm}
   \mathcal{T} _{ij}(\beta, \mu) &=& \int_0^\infty   \mathcal{T}_{ij}(\beta, \tau) P(\tau)\;  d\tau \nonumber \\
   &=& 1- \left(1+\frac{\beta}{\mu}\frac{a_i + a_j}{N} \right)^{-1} 
\simeq \frac{\beta}{\mu}\frac{a_i + a_j}{N}
\end{eqnarray}
in the limit of large $N$.  If we consider the process of infection as
equivalent to establishing a link between infected and susceptible
individuals and we compare this expression with
Eq.~(\ref{eq:conn_prob}), we can see that the SIR process can be
mapped to the creation of the integrated network in the activity
driven model up to a time $T = \beta /\mu$.  The epidemic threshold
will be given by the existence of a finite cluster of recovered
individuals, and therefore will coincide with the temporal percolation
threshold, i.e.
\begin{equation}
  \left(\frac{\beta}{\mu}\right)_c = T_p.
\end{equation}
The temporal percolation threshold given by Eq.~(\ref{eq:perc_real})
recovers the epidemic threshold obtained in
Ref.~\cite{liu_controlling_2013} using a mean-field rate equation
approach\footnote{Notice that in Ref.~\cite{liu_controlling_2013} the
  per capita infection rate $\beta' = 2 \av{a}\beta$ is used.}. A
particular benefit of this percolation mapping is the fact that it
makes accessible the calculation of explicit approximate forms for the
size of epidemic outbreaks, Eq.~(\ref{eq:S_appr}) (valid however in
certain limits), which are not easily available in mean-field
approximations \cite{2012arXiv1203.5351P,liu_controlling_2013}.

\section{Summary and Conclusions}
\label{sec:conclusions}

In this work we have studied the time evolution of the connectivity
properties of the integrated network ensuing from a sequence of
pairwise contacts between a set of fixed agents, defining a temporal
network. We have focused in particular in the onset of the giant
component in the integrated network, defined as the largest set of
connected agents that have established at least one contact up to a
fixed time $T$. The onset of the giant component takes place at some
percolation time $T_p$, which depends on the details of temporal
network dynamics. Considering in particular the recently proposed
activity driven model \cite{2012arXiv1203.5351P}, and building upon
the mapping of this temporal network on a network model with hidden
variables \cite{starnini_topological_2013}, we are able to provide
analytical expressions for the percolation time. Assuming lack of
degree correlations in the initial evolution of the integrated
network, the application of the generating function formalism
\cite{PhysRevE.64.026118} allows to obtain a explicit general form for
the temporal percolation threshold, as well as analytic asymptotic
expressions for the size of the giant component in the vicinity of the
threshold. These expressions turn out to be in good agreement with
numerical results for particular forms of the activity distribution
imposing weak degree correlations. For a skewed, power-law distributed
activity $F(a) \sim a^{-\gamma}$, the uncorrelated results are still
numerically correct for large values of $\gamma$. When $\gamma$ is
small, however, strong disagreements arise. 
Applying a
percolation formalism for correlated networks
\cite{citeulike:9984067}, we are able to obtain the analytical
threshold $T_p$.  For $\gamma >3$, the correlated threshold collapses
onto the uncorrelated result, which thus provides a very good
approximation to the exact result.  For small $\gamma<3$, the
percolation threshold as obtained by numerical simulation of large
networks is in very good agreement with the analytical 
prediction.

The study of the percolation properties of integrated temporal
networks opens in our view new interesting venues of future research,
related in particular to the properties of dynamical processes running
on top of them and to the coupling of their  different time
scales. One such application in the context of epidemic spreading is
the study of the SIR model, which we have shown can be mapped to a
temporal percolation problem in activity driven networks, thus
providing explicit forms (albeit valid in certain limits of weak
degree correlations) for the size of epidemic outbreaks in this class
of systems.

\begin{acknowledgments}
  We acknowledge financial support from the Spanish MICINN, under
  project No. FIS2010-21781-C02-01. R.P.-S. acknowledges additional
  financial support from ICREA Academia, funded by the Generalitat de
  Catalunya
\end{acknowledgments}


\bibliographystyle{apsrev4-1}


\bibliography{Bibliography.bib}

\end{document}